# A Catalog of Spectra, Albedos, and Colors of Solar System Bodies for Exoplanet Comparison

J. H. Madden[1,2] & Lisa Kaltenegger[1,2]

[1]Carl Sagan Institute, [2]Astronomy Department Cornell University, 311 Space Science Building, Ithaca, NY 14850, jmadden@astro.cornell.edu

ABSTRACT

We present a catalog of spectra and geometric albedos, representative of the different types of Solar System bodies, from 0.45 to 2.5 microns. We analyzed published calibrated, un-calibrated spectra, and albedos for Solar System objects and derived a set of reference spectra and reference albedo for 19 objects that are representative of the diversity of bodies in our Solar System. We also identified previously published data that appears contaminated. Our catalog provides a baseline for comparison of exoplanet observations to 19 bodies in our own Solar System, which can assist in the prioritization of exoplanets for time intensive follow-up with next generation Extremely Large Telescopes (ELTs) and space based direct observation missions. Using high and low-resolution spectra of these Solar System objects, we also derive colors for these bodies and explore how a color-color diagram could be used to initially distinguish between rocky, icy, and gaseous exoplanets. We explore how the colors of Solar System analog bodies would change when orbiting different host stars. This catalog of Solar System reference spectra and albedos is available for download through the Carl Sagan Institute.

1 INTRODUCTION

The first spectra of extrasolar planets have already been observed for gaseous bodies (e.g. Dyudina et al., 2016; Kreidberg et al., 2014; Mesa et al., 2016; Sing et al., 2016; Snellen et al., 2010). To aid in comparative planetology exoplanet observations will require an accurate set of disk-integrated reference spectra, and albedos of Solar System objects. To establish this catalog for the solar system we use disk-integrated spectra from several sources. We use un-calibrated and calibrated spectra as well as albedos when available from the literature to compile our reference catalog. About half of the spectra and albedos we derive in this paper are based on un-calibrated observations obtained from the Tohoku-Hiroshima-Nagoya Planetary Spectral Library (THN-PSL)(Lundock et al., 2009), which provides a large coherent dataset of un-calibrated data taken with the same telescope. Our analysis shows contamination of part of that dataset, as discussed in section 2.1 and 2.2, therefore we only include a subset of their data in our catalog (see discussion 4.3).

This paper provides the first catalog of calibrated spectra (Fig. 1) and geometric albedos (Fig. 2) of 19 bodies in our Solar System, representative of a wide range of object types: all 8 planets, 9 moons (representing, icy, rocky, and gaseous moons), and 2 dwarf planets (Ceres in the Asteroid belt and Pluto in the Kuiper belt). This catalog is available through the Carl Sagan Institute[1] to enable comparative planetology beyond our Solar System.

---

[1] www.carlsaganinstitute.org/data/

Several teams have shown that photometric colors of planetary bodies can be used to initially distinguish between icy, rocky, and gaseous surface types (Krissansen-Totton et al., 2016; Cahoy et al., 2010; Lundock et al., 2009; Traub, 2003) and that models of habitable worlds lie in a certain color space (Krissansen-Totton et al., 2016; Hegde & Kaltenegger, 2013; Traub, 2003). We expand these earlier analyses from a smaller sample of Solar System objects to 19 Solar System bodies in our catalog, which represent the diversity of bodies in our Solar System. In addition, we explore the influences of spectral resolution on characterization of planets in a color-color diagram by creating low resolution versions of our data. Using the derived albedos, we also explore how colors of analog planets would change if they were orbiting other host stars. Section 2 of this paper describes our methods to identify contamination in the THN-PSL data, derive calibrated spectra, albedos, and colors from the un-calibrated THN-PSL data and how we model the colors of the objects around the Sun and other host stars. Section 3 presents our results, Section 4 discusses our catalog, and Section 5 summarizes our findings.

2 METHOD

We first discuss our analysis of the THN-PSL data and how we identified contaminated data in detail, then discuss how we derived spectra and albedo from the uncontaminated data. Finally we discuss spectra and albedo from other data sources for our catalog.

*2.1 Calibrating the Spectra of Solar System Bodies from the THN-PSL*

The THN-PSL is a collection of observations of 38 spectra for 18 Solar System objects observed over the course of several months in 2008. The spectra of one of the objects, Callisto, was contaminated and could not be re-observed, while the spectrum for Pluto in the database is a composite spectrum of both Pluto and Charon. We analyzed the data for the 16 remaining Solar System objects for additional contamination and found 6 apparently contaminated objects among them, leaving 10 objects in the database that do not appear contaminated. Their albedos are similar to published values in the literature for the wavelength range such data is available for. We show the derived albedos for both contaminated and uncontaminated data from the database in Fig. 3, compared to available values from the literature for these bodies.

The THN-PSL data were taken in 2008 using the TRISPEC instrument while on the Kanata Telescope at the Higashi-Hiroshima observatory. TRISPEC (Watanabe et al., 2005) splits light into one visible channel and two near-infrared channels giving a wavelength range of 0.45−2.5μm. The optical band covered 0.45-0.9μm and had a resolution of $R = \lambda/\Delta\lambda = 138$. The first IR channel has a coverage from 0.9-1.85μm and had a resolution of $R = 142$. The second IR channel has a coverage from 1.85-2.5μm and had a resolution of $R = 360$. Note that the slit subtends 4.5 arcseconds by 7 arcmin meaning that spectra for larger bodies such as Saturn and Jupiter were not disk integrated (see discussion).

As discussed in the original paper, all spectra are unreliable below 0.47μm and between 0.9-1.0μm from a dichroic coating problem with the beam splitters. Near 1.4μm and 1.8μm the Earth's water absorption degrades the quality and beyond 2.4μm thermal contamination is an issue. These wavelength regions are grayed out in all relevant figures in our paper but do not influence our color analysis, due to the choice of filters. The raw data available for download includes all data points. The THN-PSL paper discusses several initial observations of moons that were contaminated with light from their host planet rendering their spectra inaccurate (080505 Callisto, 081125 Dione, 080506 Io, and 080506 Rhea). These objects (with the exception of Callisto) were observed again and the extra light was removed in a different fashion to more accurately correct the spectra (Lundock et al., 2009). Callisto was not re-observed and therefore the THN-PSL Calisto data remained contaminated (Fig.3).

The fluxes of the published THN-PSL observations were not calibrated but arbitrary normalized to the value of 1 at 0.7μm. This makes the dataset generally useful to compare the colors of the uncontaminated objects, as shown in the original paper, but limits the data's usefulness as reference for extrasolar planet observations because geometric albedos can only be derived from calibrated spectra. The conversion factors used in the original publication were not available (Ramsey Lundock, private communication).

However, in addition to the V magnitude, the THN-PSL gives the color differences: V-R, R-I, R-J, J-K, and H-Ks for each observation, providing the R, I, and J magnitudes. Therefore, we used the published V, R, I, and J magnitudes to derive the conversion factor for each spectrum to match the published color magnitudes and to calibrate the THN-PSL observations.

We define the conversion factor $k$ such that $kf_{norm} = f$ where $f_{norm}$ and $f$ are the normalized and absolute spectra respectively. Adapting the method outlined in Fukugita et al. (1995) the magnitude in a single band using the filter response, $V$, and the spectrum of Vega, $f_{vega}$, is given by equation (1). The spectrum of Vega, (Bohlin, 2014)[2] as well as the filter responses, are the same as in the THN-PSL publication and shown in Fig. 4 and Fig. 5 respectively. The filters we used are V (Johnson & Morgan, 1953); R and I (Bessell, 1979; Cousins, 1976); J, H, K, and Ks (Tokunaga et al., 2002). Since the THN-PSL paper recorded the V, R, I, and J color magnitudes for each object we derive the conversion factor to obtain each magnitude and average them to obtain $k$. For example, we substitute $k_V f_{norm} = f$ in equation (2) and isolate $k_V$ as shown in equation (3) to calculate the conversion factors for the V band. The conversion factors for each band for a single object were averaged and used to calibrate the normalized spectra.

$$m_V = -2.5 \, log_{10} \left( \frac{\int V f d\lambda}{\int V f_{vega} d\lambda} \right) \quad (1)$$

$$m_V = -2.5 \, log_{10} \left( \frac{\int V k_V f_{norm} d\lambda}{\int V f_{vega} d\lambda} \right) \quad (2)$$

$$k_V = \left( \frac{\int V f_{vega} d\lambda}{\int V f_{norm} d\lambda} \right) 10^{\frac{-m_V}{2.5}} \quad (3)$$

$$k = \left( k_V + k_R + k_I + k_J \right)/4 \quad (4)$$

We used this method to calibrate the THN-PSL data for each object. When comparing the coefficient of variation (CV) of the conversion factors for each body we found that the data showed two distinct groups, one with a CV greater than 14% and another with a CV smaller than 6%. We use that distinction to set the level of the conversion factor for uncontaminated spectra to (CV>6%) over the different filter bins. If the CV value was in the second group (CV>14%), the data is flagged as contaminated and not used in our catalog. The nature of this contamination is unclear, it could be photometric error during the observation, excess light from the host planet or other effects that influenced the observations. The values calculated for the $k_V, k_R, k_I, k_J, k$, and the CV for each observation is given in Table A1.

---

[2] www.stsci.edu/hst/observatory/crds/calspec.html (alpha_lyr_stis_008.fits)

*2.2 Albedos of Solar System Bodies*

We then derive the geometric albedo from the calibrated spectra as a second part of our analysis (see Table 1 and Table 2 for references) by dividing the observed flux of the Solar System bodies by the solar flux and accounting for the observation geometry as given in equation (5) (de Vaucouleurs, 1964).

$$p = \frac{d^2 a_b^2 f}{\phi(\alpha) R_b^2 a_\oplus^2 f_{sun}} \qquad (5)$$

where $d$ is the separation between Earth and the body, $a_b$ the distance between the Sun and the body at the time of observation, and $a_\oplus$ the semi major axis of Earth. $f_{sun}$ and $f$ are the fluxes from the Sun seen from Earth and the body seen from Earth respectively, $R_b$ is the radius of the body being observed, and $\phi(\alpha)$ is the value of the phase function at the point in time the observation was taken. For $f_{sun}$, we used the standard STIS Sun spectrum (Bohlin et al., 2001)[3] shown in Fig. 4. If the geometric albedo exceeds 1, the data is flagged as contaminated and not used in our catalog.

Note that we also compared the spectra that were flagged as contaminated in this 2-step analysis with the available data and models from other groups (Fig. 3). All flagged spectra show a strong difference in albedo for these bodies observed by other teams, supporting our analysis method (see Fig. 3).

*2.3 Using colors to characterize planets*

We use a standard astronomy tool, a color-color diagram, to analyze if we can distinguish Solar System bodies based on their colors and what effect resolution and filter choice has on this analysis. Several teams have shown that photometric colors of planetary bodies can be used to initially distinguish between icy, rocky, and gaseous surface types (Krissansen-Totton et al., 2016; Cahoy et al., 2010; Lundock et al., 2009; Traub, 2003). We calculated the colors from high and low-resolution spectra to mimic early results from exoplanet observations as well as explored the effect of spectral resolution on the colors and their interpretation. The error for colors derived from the THN-PSL data was calculated by adding the errors used by Lundock et al., 2009 and the error accumulated through the conversion process of 6% in the $k$ value. This gives $\Delta(J-K) = \pm 0.34$ and $\Delta(R-J) = \pm 0.28$ for the error values. We reduce the high-resolution data of $R = 138 - 360$ to $R = 8$ in order to mimic colors that are generated from low-resolution spectra as shown in Fig. 6. The colors at high resolutions were used to determine the best color-color combination for surface and atmospheric characterization, a process that was repeated for colors derived from low resolution spectra.

We also explored how to characterize Solar System analog planets around other host stars using their colors by placing the bodies at an equivalent orbital distance around different host stars (F0V, G0V, M0V, and M9V). We used stellar spectra for the host stars from the Castelli and Kurucz Atlas (Castelli & Kurucz, 2004)[4] and the PHOENIX library (Husser et al., 2013)[5] (Fig. 4). As a first order approximation, we have assumed that the albedo of the object would not change under this new incoming stellar flux (See discussion).

---

[3] www.stsci.edu/hst/observatory/crds/calspec.html (sun_reference_stis_002.fits)

[4] www.stsci.edu/hst/observatory/crds/castelli kurucz atlas.html (F0V, G0V, K0V, M0V)

[5] http://phoenix.astro.physik.uni-goettingen.de (M9V)

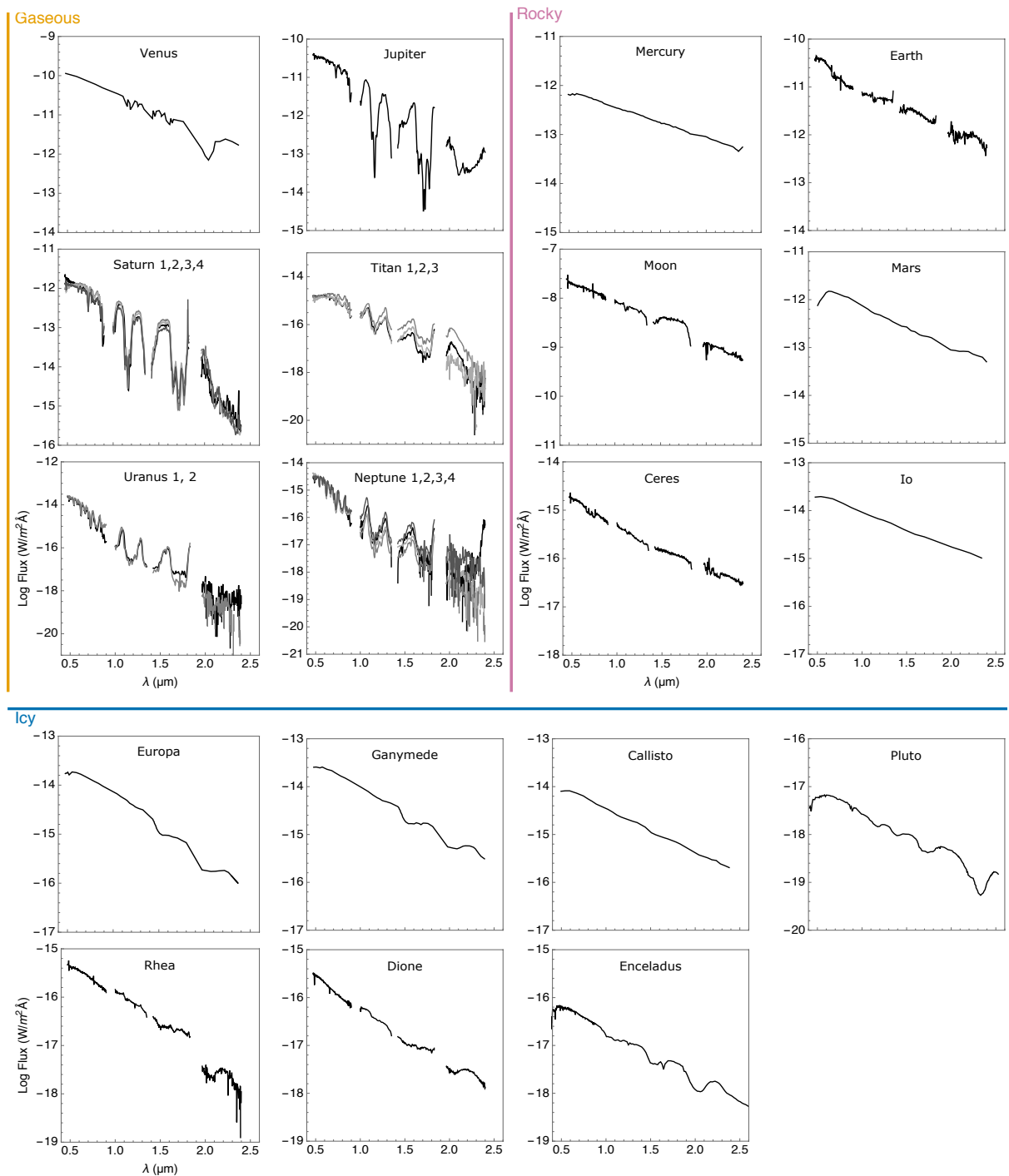

**FIG. 1.** Spectra for 19 Solar System bodies for Ceres, Dione, Earth, Jupiter, Moon, Neptune, Rhea, Saturn, Titan, Uranus (albedos calculated in this paper based on un-calibrated data by Lundock et al., 2009), Callisto (Spencer, 1987), Enceladus (Filacchione et al., 2012), Europa (Spencer, 1987), Ganymede (Spencer, 1987), Io (Fanale et al., 1974), Mars (McCord & Westphal, 1971), Mercury (Mallama, 2017), Pluto (Lorenzi et al., 2016; Protopapa et al., 2008), and Venus (Meadows 2006 (theoretical); Pollack et al. 1978 (observation)). Items are arranged by body type then by distance from the Sun.

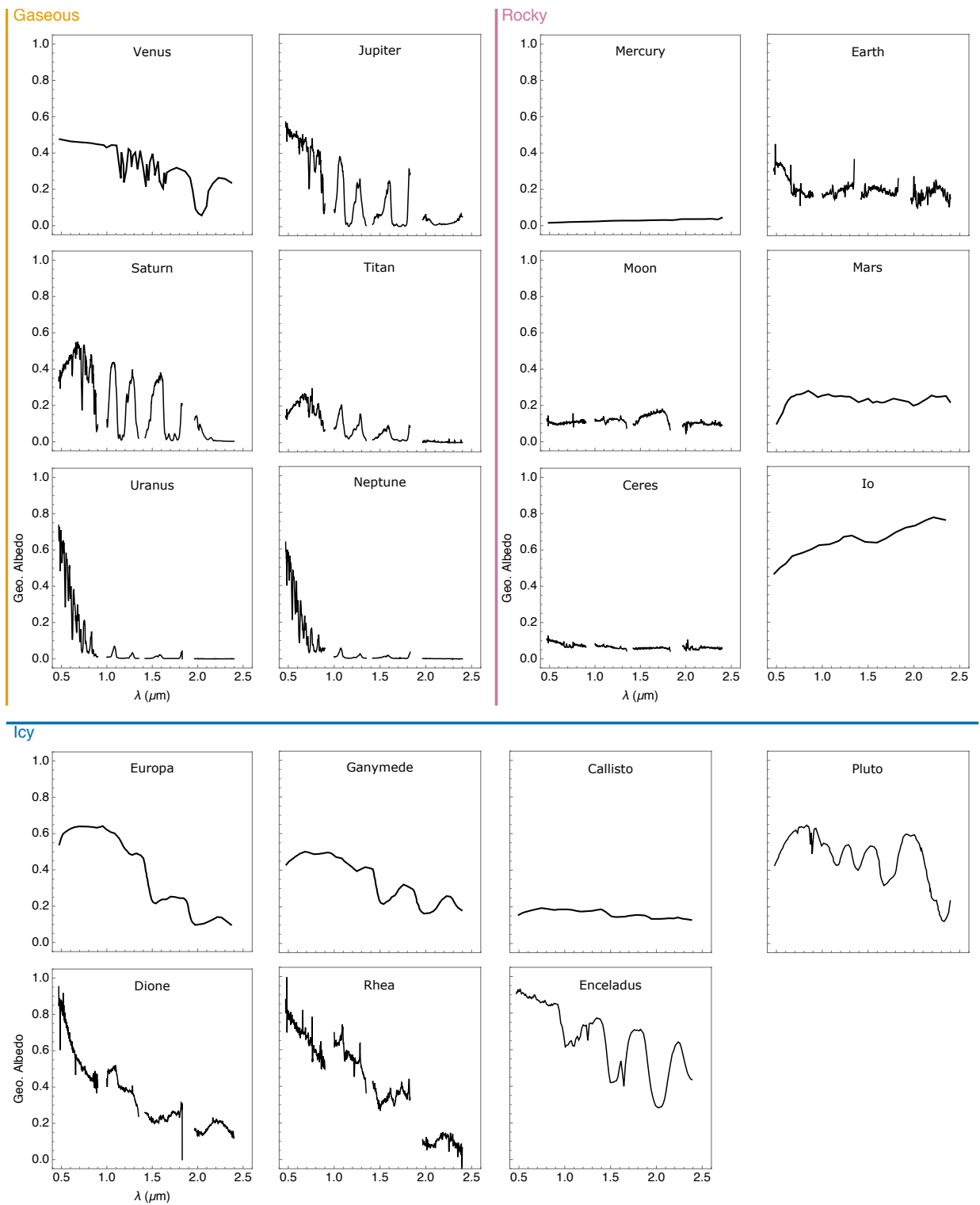

**FIG.2.** Geometric albedos for 19 Solar System bodies for Ceres, Dione, Earth, Jupiter, Moon, Neptune, Rhea, Saturn, Titan, Uranus (albedos calculated in this paper based on un-calibrated data by Lundock et al., 2009), Callisto (Spencer, 1987), Enceladus (Filacchione et al., 2012), Europa (Spencer, 1987), Ganymede (Spencer, 1987), Io (Fanale et al., 1974), Mars (McCord & Westphal, 1971), Mercury (Mallama, 2017), Pluto (Lorenzi et al., 2016; Protopapa et al., 2008), and Venus (Meadows 2006 (theoretical); Pollack et al. 1978 (observation)). Items are arranged by body type then by distance from the Sun.

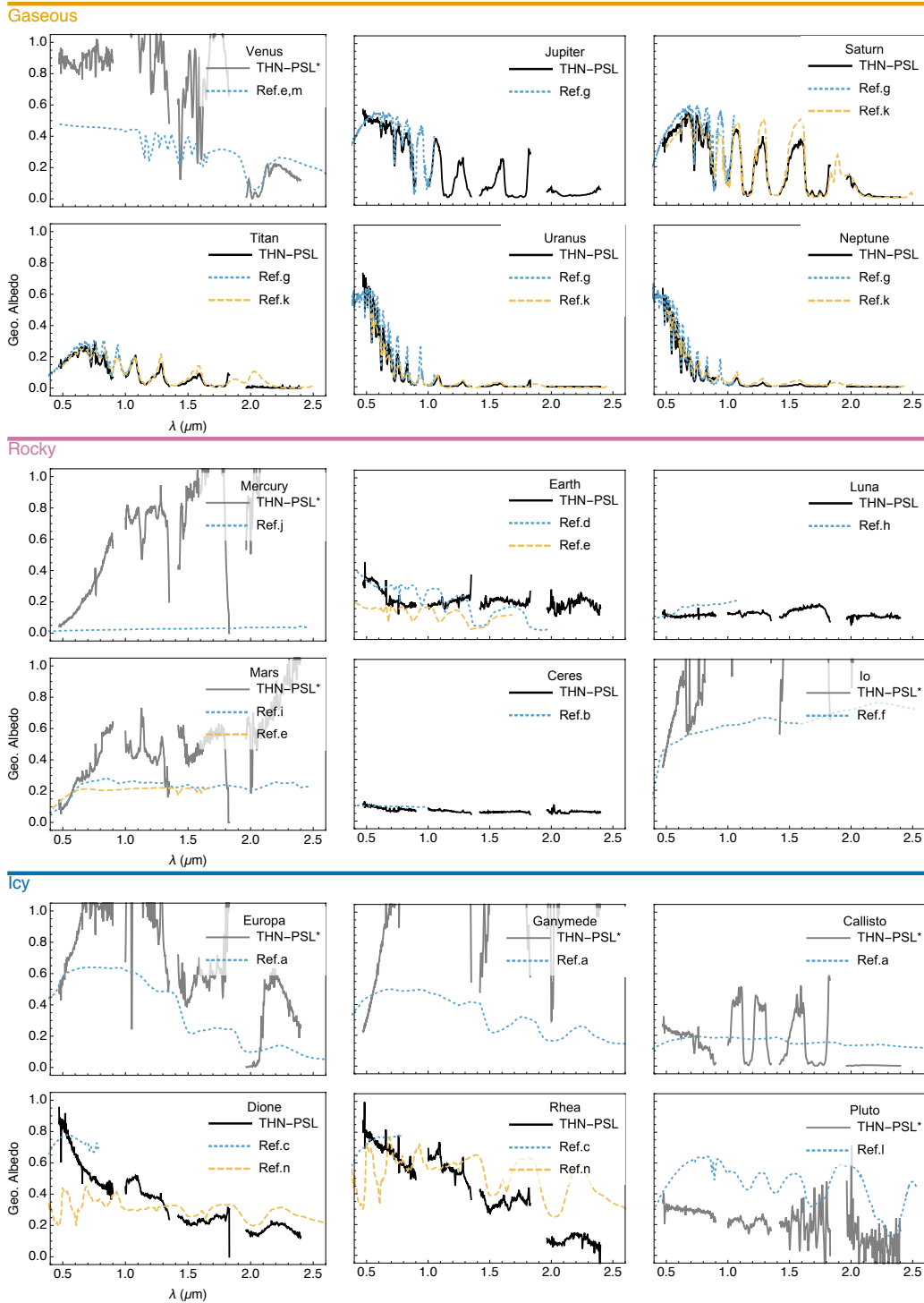

**FIG.3.** A comparison of geometric albedos for the Solar System bodies in our catalog between published values and the albedo calculated from the THN-PSL data. THN-PSL data based albedos are denoted with solid lines for uncontaminated data and with an asterisk and gray line if contaminated. References for comparison albedos: [a](Spencer, 1987), [b](Reddy et al., 2015), [c](Noll et al., 1997), [d](Kaltenegger et al., 2010), [e](Meadows, 2006), [f](Fanale et al., 1974), [g](Karkoschka, 1998), [h](Lane & Irvine, 1973), [i](McCord & Westphal, 1971), [j](Mallama, 2017), [k](Fink & Larson, 1979), [l](Lorenzi et al., 2016; Protopapa et al., 2008), [m](Pollack et al., 1978), [n](Cassini VIMS – NASA PDS). Items are arranged by body type then by distance from the Sun.

## 3 RESULTS

### 3.1 A spectra and albedo catalog of a diverse set of Solar System Objects

We assembled a reference catalog of 19 bodies in our Solar System as a baseline for comparison to upcoming exoplanet observations. To provide a wide range of Solar System bodies in our catalog we compiled and analyzed data from un-calibrated and calibrated spectra of previously published disk-integrated observations.

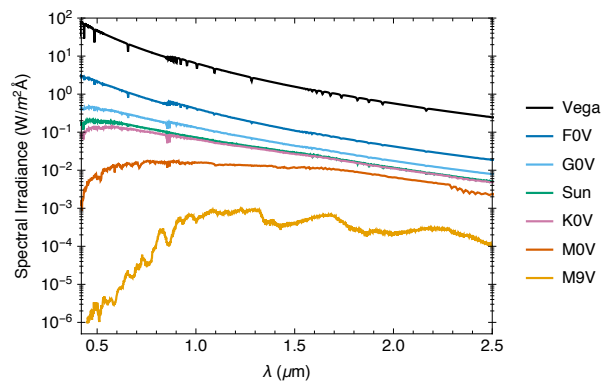

**FIG.4.** Reference spectra used for calibration (Sun and Vega) and model spectra used for host stars at 1 AU (F0V, G0V, K0V, M0V, M9V). Vega was multiplied by $10^{13}$ to fit on the same plot.

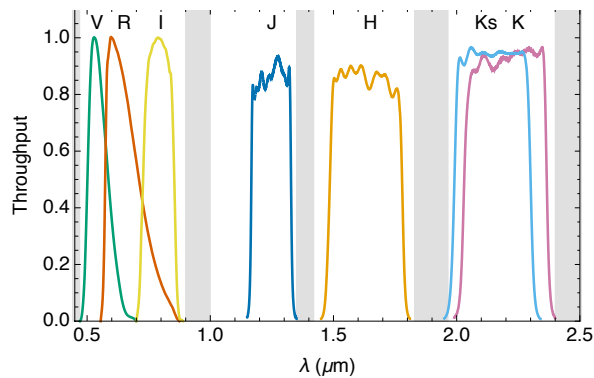

**FIG.5.** Standard filters used for flux calibration and color calculations. Gray bands show the wavelength range where the observed fluxes from the THN-PSL are not reliable.

Our catalog contains spectra and geometric albedo of the 8 planets: Mercury (Mallama, 2017), Venus (Meadows, 2006; Pollack et al., 1978), Earth (Lundock et al., 2009), Mars (McCord & Westphal, 1971), Jupiter, Saturn, Uranus, and Neptune (Lundock et al., 2009). 9 moons: Io (Fanale et al., 1974), Callisto, Europa, Ganymede (Spencer, 1987), Enceladus[6] (Filacchione et al., 2012), Dione, Rhea, the Moon, and Titan (Lundock et al., 2009), and 2 dwarf planets: Ceres (Lundock et al., 2009) and Pluto (Lorenzi et al., 2016; Protopapa et al., 2008).

For the 8 planets of the Solar System, 9 moons (Callisto, Dione, Europa, Ganymede, Io, the Moon, Rhea, Titan), and 2 dwarf planets (Ceres and Pluto) we present the absolute fluxes in Fig.1 and the geometric albedos in Fig. 2.

### 3.2 Contaminated spectra in the THN-PSL dataset

When we derived the geometric albedo from the calibrated THN-PSL spectra as the second part of our analysis, we found that 6 objects (Io, Europa, Ganymede, Mercury, Mars, and Venus) display geometric albedos exceeding 1, indicating that the measurements are contaminated (see Table 2, Fig. 3). We compared the albedo of these six observations to previously published values in the literature (Mallama, 2017; Meadows, 2006; Mallama et al., 2002; Spencer et al., 1995; Spencer, 1987; Buratti & Veverka, 1983; Pollack et al., 1978; Fanale et al., 1974; McCord & Westphal, 1971) and found substantial differences over the wavelength covered by the different teams (Fig.3). We list the 7 bodies with contaminated THN-PSL measurements in Table 2.

### 3.3 Spectra not flagged as contaminated in the THN-PSL dataset

Table 1 lists the spectra of the 10 bodies from the THN-PSL database, which were not flagged as contaminated and are part of our catalog, as well as the Pluto-Charon spectrum. It shows the properties we used to calculate their albedos, once we un-normalized the un-calibrated data as well as references to previously published albedos. Note that we did not use the THN-PSL Pluto-Charon spectrum in our analysis because it is not a Pluto spectrum. Instead we use the spectrum for Pluto published by two teams (Lorenzi et al., 2016; Protopapa et al., 2008) that cover the wavelength range requires for our analysis. We show both spectra in Fig.3 for completeness.

---

[6] Data available on NASA's Planetary Data Archive: (v1640517972_1, v1640518173_1, v1640518374_1)

We compared the derived albedo of the 10 bodies from the THN-PSL database, which were not flagged as contaminated, against disk-integrated spectra and albedo from observations or models in the literature for the wavelengths available. Our derived albedos are in qualitative agreement with previously published data (Fig. 3) for Ceres (Reddy et al., 2015); Dione and Rhea (Noll et al., 1997; Cassini VIMS); Earth (Kaltenegger et al., 2010; Meadows, 2006); the Moon (Lane & Irvine, 1973); Jupiter, Saturn, Uranus, Neptune, and Titan (Karkoschka, 1998; Fink & Larson, 1979). We simulated their absolute fluxes with the same observation geometry as the THN-PSL spectra to be able to compare them (Fig. 3). Note that small changes are likely due to observation geometry as well as the changes in the atmospheres over the time between observations. Giant planets have daily variations in brightness (Belton et al., 1981). For completeness we include the THN-PSL observation of the combined spectrum of Pluto and Charon and compare it to the albedo of Pluto (Lorenzi et al., 2016; Protopapa et al., 2008). We averaged several Cassini VIMS observations together and used them as references for Rhea and Dione[7].

*3.4 Using Color-color diagrams to initially characterize Solar System bodies*

To qualify the Solar System objects in terms of extrasolar planet observables, we consider whether they are gaseous, icy, or rocky bodies and do not distinguish between moons and planets. Thus, Titan and Venus are both gaseous bodies in our analysis since only their atmosphere is being observed at this wavelength range.

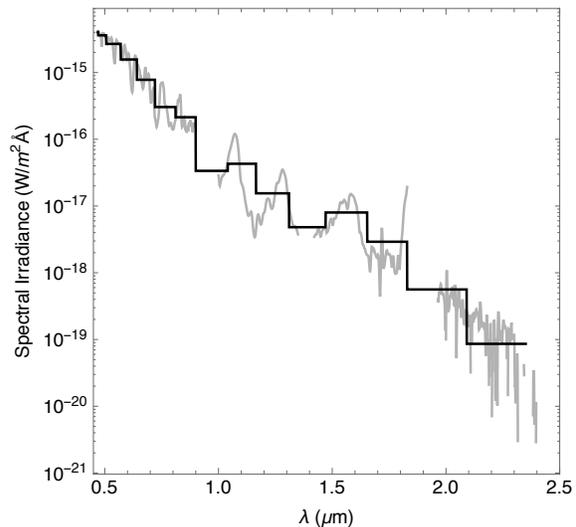

**FIG.6.** An example of a reduced resolution spectrum compared to its high-resolution observations.

Fig. 7 shows the spectra as well as the colors for the three subcategories in our catalog. The top panel shows gaseous bodies: Jupiter, Saturn, Uranus, Neptune, Venus, Titan. The middle panel shows rocky bodies: Mars, Mercury, Io, Ceres, Earth, and the Moon. The bottom panel shows icy bodies: Ganymede, Dione, Rhea, Callisto, Pluto, Europa, and Enceladus. Each surface type occupies its own color space in the diagram. To explore how the resolution of the available spectra and thus the observation time available would influence this classification, we reduced the spectral resolution for all spectra to $R = \lambda/\Delta\lambda$ of 8.

---

[7] Data available on NASA's PDS:
Rhea – v1498350281_1, v1579258039_1, v1579259351_1
Dione - v1549192526_1, v1549192731_1, v1549193961_1

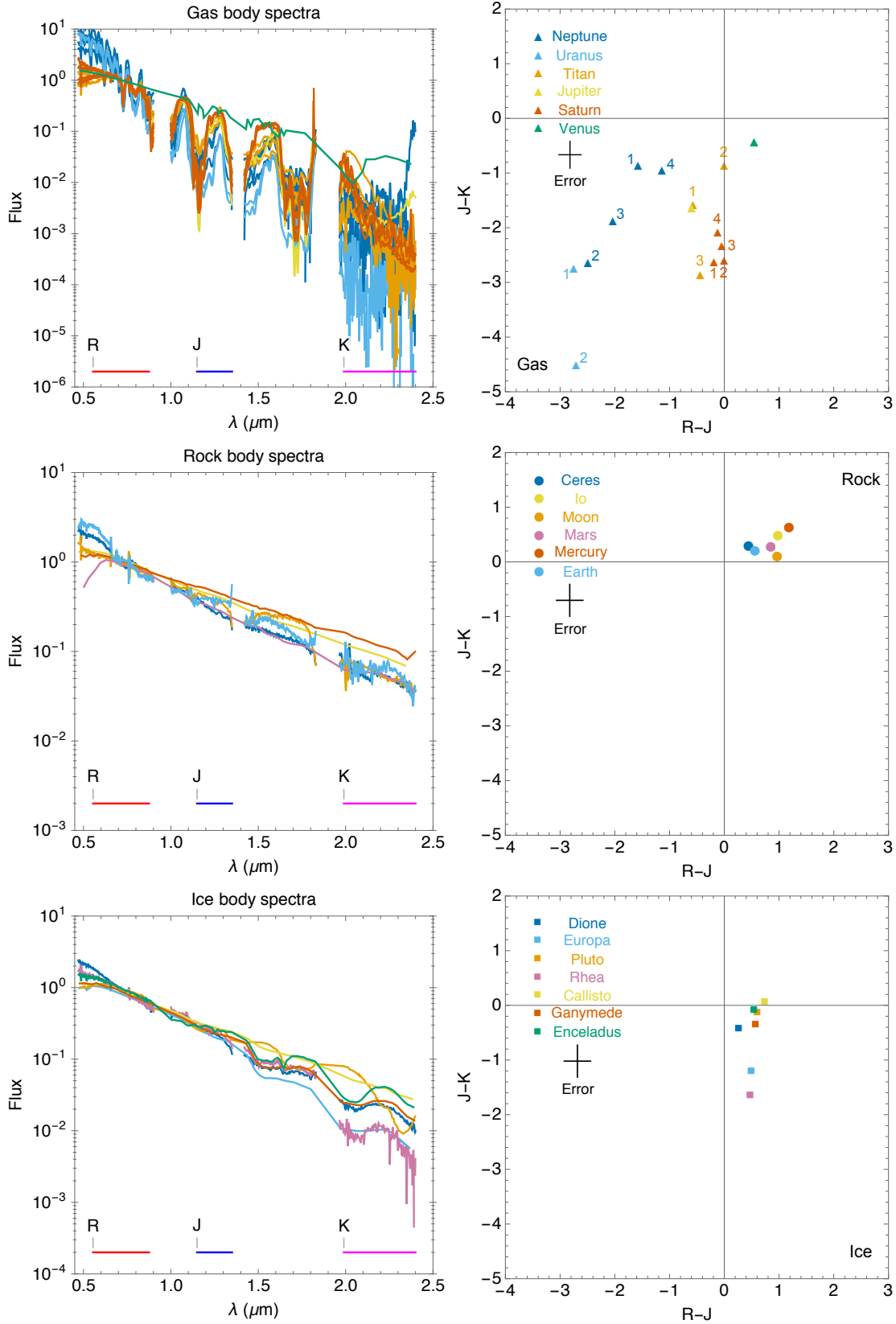

**FIG.7.** Spectra and color-color diagrams for gaseous, rocky and icy bodies of the Solar System. Previously published data was used for bodies that were contaminated in the THN-PSL following the references in Fig. A1.

We find that the derived colors of the Solar System bodies do not shift substantially (Fig. 8), showing that colors derived from high and low-resolution spectra provide similar capabilities for first order color-characterization of a Solar System object. While a slight shift occurs in the color-color diagram, the three different Solar System surface types (gaseous, rocky and icy) can still be distinguished, showing that colors from low resolution spectra can be used for first order characterization of bodies in our Solar System.

We chose a lower resolution of $R = 8$ since the bin width near the K-band becomes larger than the K-band filter itself at lower resolutions. Bandwidth is directly proportional to the amount of light collected by a telescope and thus the time needed for observation.

If low-resolution spectra could initially characterize a planet, exoplanets could be prioritized for time-intense high-resolution follow-up observations from their colors. At a lower resolution, a higher signal to noise ratio is required to achieve the same distinguishability as an observation at high resolution. The ratio of the integral uncertainties of two spectra at different resolutions, $\Delta I_A$ and $\Delta I_B$, is proportional to the number of bins being integrated over, $N$, and the measurement uncertainty of each bin, $\delta m$, as shown in equation 6.

$$\frac{\Delta I_A}{\Delta I_B} \propto \sqrt{\frac{N_B}{N_A}} \frac{\delta m_A}{\delta m_B} \qquad (6)$$

We explored different filter combinations to best distinguish between icy, gaseous and rocky bodies. We find that R-J versus J-K colors distinguish the bodies best, (see also Krissansen-Totton et al., 2016; Cahoy et al., 2010; Lundock et al., 2009; Traub, 2003).

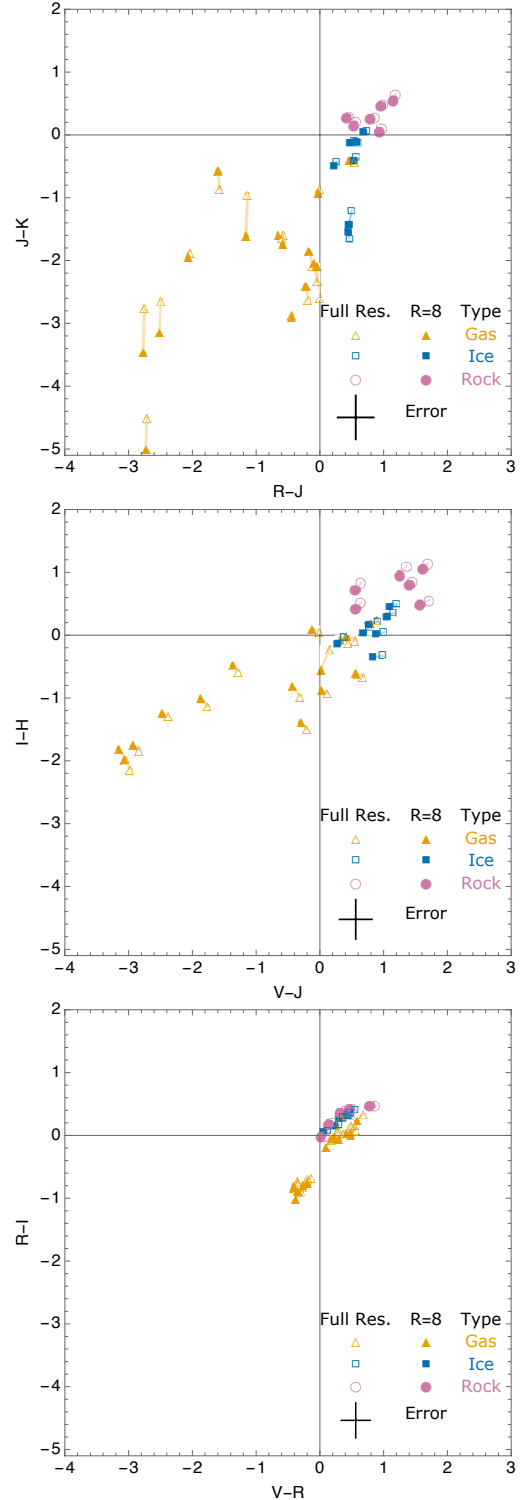

**FIG.8.** Comparison of colors calculated using low (filled symbols, $R = 8$) versus high resolution (non-filled symbols, $R = 138$, 142 and 360 for THN-PSL data) spectra for 19 Solar System bodies around the Sun. Each panel shows a different filter combination and the symbols represent the three surface types; gaseous (square), icy (triangle) and rocky (circle).

If only a smaller wavelength range is available, such as V through H or V through I, Fig. 8 shows which alternate filter combinations can still separate the surface types. However, Fig. 8 shows that a wider wavelength range improves the characterization of surfaces for Solar System objects substantially. The success of using this method to characterize the Solar System reduces with narrower wavelength coverage. Long wavelengths (J and K band) especially help distinguish different kind of Solar System bodies (Fig 8). To characterize all bodies in the Solar System it is important to have wavelength coverage of the visible and near IR at a resolution that distinguishes each band.

*3.5 Colors of Solar System analog bodies orbiting different host stars*

To provide observers with the color-space where Solar System analog exoplanets could be found, we use the albedos shown in Fig. 2 to explore the colors of similar bodies orbiting different host stars. For airless bodies the albedo is a direct surface measurement, therefore that assumption should be valid for similar surface composition. For objects with substantial atmospheres that can be influenced by stellar radiation, individual models are needed to assess whether the albedo of a system's bodies would notably change due to the different host star flux. Note that Earth's albedo would not change significantly from F0V to M9V host stars in the wavelength range considered here (see Rugheimer et al., 2013, 2015).

Fig. 9 shows the colors of the Solar System analog bodies orbiting other host stars. Because their albedo is assumed to be constant the shift closely mimics the shift in colors of the host star. For hotter host stars the colors shift to a bluer section of the color-color diagram (F0V). For cooler host stars the colors shift toward a redder portion of the color space (M0V, and M9V). This provides insights for observers into where the divisions in color-space of rocky, icy, and gaseous bodies lie depending on the host star's spectral class.

4 DISCUSSION
*4.1 Change in colors of Gaseous Planets*

Some gaseous bodies in the THN-PSL with multiple spectra (Uranus, Neptune, and Titan) show variations in their colors larger than the error (Fig. 7). Gaseous bodies are known to vary in brightness over timescales shorter than the time between these observations (Belton et al., 1981), consistent with the THN-PSL data. This indicates that any sub-divide for gaseous bodies would be challenging from their colors alone. However the K-band is also more susceptible to photometric error as discussed in the THN-PSL paper (Lundock et al., 2009), which could add to the observed differences. Multiple uncontaminated observations across the same wavelengths for rocky or icy bodies are not available in the literature, therefore we cannot assess whether a spread in colors also exists for rocky or icy bodies, independent of viewing geometry.

*4.2 Non disk-integrated spectra of some objects*

Due to the finite field of view of the TRISPEC instrument the observations of the Earth, Moon, Jupiter, and Saturn were not disk integrated. A disk integrated spectrum is preferred because it averages the light from the entire body instead of from a small region of its surface. The spectrograph slit was centered on the planet and aligned longitudinally for Jupiter and Saturn making the spectra as representative of the entire surface as possible. When comparing their spectra to other sources, the spectra shows a good match to disk integrated spectra (Karkoschka, 1998; Fink & Larson, 1979). This could not be done for the Moon and the Earthshine observations leading to variations in their spectra from previously published data.

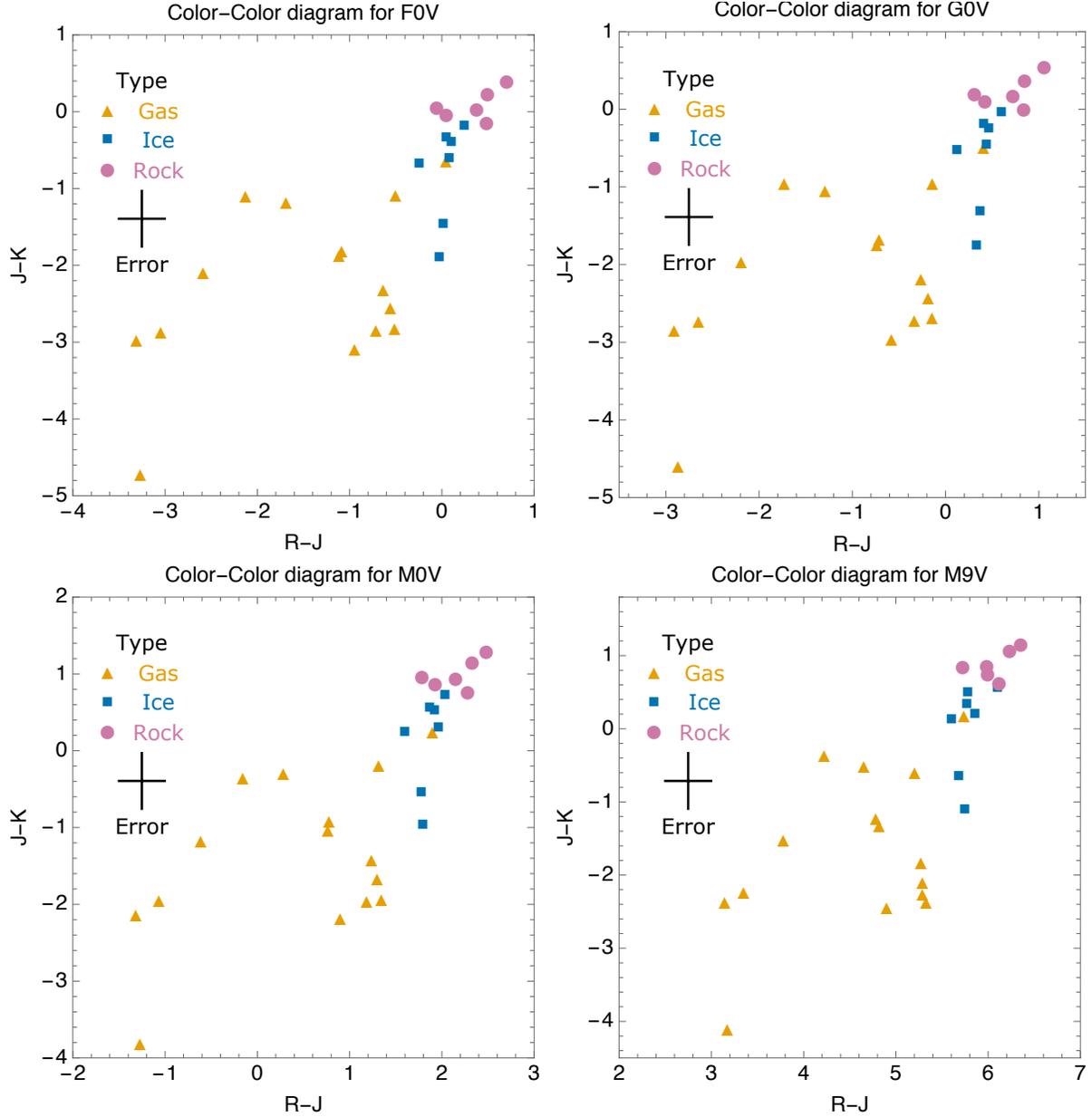

**FIG.9.** Colors of Solar System bodies around different host stars. Here we show the colors of Solar System bodies for an F0V (upper-left), a G0V (upper-right), an M0V (lower-left), and an M9V (lower-right) host star.

*4.3. Spectra derived from the THN-PSL dataset*

We have used several spectra of planets and moons from the THN-PSL dataset that did not appear contaminated in our analysis. The contamination of 6 objects in the THN-PSL database raises questions about the viability of the spectra in this database in general.

We compared the 10 bodies that we used in our catalog, which were not flagged as contaminated against disk-integrated spectra and albedo from observations or models in the literature. These observations or models were not available for the whole wavelength range, thus we could not compare the full wavelength range, however the range covered shows qualitative agreement with previously published data (Fig. 3) and thus we have included the spectra and albedos we derived from the un-calibrated THN-PSL data in our analysis. For Earth and the Moon time variability of the spectra can be explained because observations of the Earth and the Moon were not disk integrated, due to the spectrographic slit as discussed in 4.2. Note that for most solar system objects, reliable disc-averaged spectra for different times are lacking, which are observations that would be useful for future exoplanet comparisons.

*4.4 Similarity of the color of water and rock*

The primarily liquid water surface of the Earth is unique in the Solar System however, this is not apparent in the color-color diagrams (Fig 7, 8, and 9). This is because water and rock share a similar, relatively flat, albedo over the 0.5-2.5μm wavelength range. This specific color-color degeneracy for rock and water can be broken if shorter wavelength observations are available (see also Krissansen-Totton et al., 2016).

*4.5 Color of $CO_2$ atmospheres appear similar to icy surfaces*

Venus has the interesting position of being a rocky planet that has a gaseous appearance but lies amongst the colors of the icy bodies in the color-color diagrams. This is due to Venus having a primarily $CO_2$ atmosphere which provides a similarly sloped albedo as ice in this wavelength range. This shows the limits of initial characterization through a color-color diagram. It will make habitability assessments from colors alone of terrestrial planets especially on the edges of their habitable zones very difficult since $CO_2$ is likely to be present. Estimates of the effective stellar flux that reaches the planet or moon could help to disentangle the ice/$CO_2$ degeneracy on the inner edge of the Habitable Zone. On the outer edge of the Habitable Zone both surface types should be present, $CO_2$-rich atmospheres as well as icy bodies, therefore higher resolution spectra will be needed to break such degeneracy.

*4.6 Spotting the absence of methane in a gas planet's colors*

The absence of methane in Venus' atmosphere makes it distinguishable from the other gaseous objects in our Solar System in the color-color diagrams. More information about the atmospheric composition of exoplanets and exomoons would be needed before we can assess whether we could derive similar inferences for other planetary systems.

*4.7 Colors of objects that are made of 'dirty snow'*

In the color-color diagrams, Ganymede and Callisto fall in the region between rocky and ice bodies due to their high amount of 'dirty snow' compared to the other bodies in the icy body category. Given the error bars in their colors, these two bodies could be placed in either the rocky or icy categories. Such rocky-icy bodies are anticipated in other planetary systems as well and should lie in the color space between the icy and rocky bodies like in our Solar System.

5 CONCLUSIONS

We present a catalog of spectra, and geometric albedos for 19 Solar System bodies, which are representative of the types of surfaces found throughout the Solar System for wavelengths from 0.45-2.5 microns. This catalog provides a baseline for comparison of exoplanet observations to the most closely studied bodies in our Solar System. The data used and created by this paper is available for download through the Carl Sagan Institute[8].

We show the utility of a color-color diagram to distinguish between rocky, icy, and gaseous bodies in our Solar System for colors derived from high as well as low-resolution spectra (Fig. 7 and 8) and initially characterize extrasolar planets and moons. The spectra, albedo and colors presented in this catalog can be used to prioritize time-intensive follow up spectral observations of extrasolar planets and moons with current and next generation like the Extremely Large Telescopes (ELTs). Assuming an unchanged albedo, Solar System body analog exoplanets shift their position in a color-color diagram following the color change of the host stars (Fig. 9). Detailed spectroscopic characterization will be necessary to confirm the provisional categorization from the broadband photometry suggested here, which is only based on planets and moons of our own Solar System.

---

[8] www.carlsaganinstitute.org/data/

Planetary science broke new ground in the 70s and 80s with spectral measurements for Solar System bodies. Exoplanet science will see a similar renaissance in the near future, when we will be able to compare spectra of a wide range of exoplanets to the catalog of bodies in our Solar System.

ACKNOWLEDGMENTS
We Thank Gianrico Filacchione, Ramsey Lundock, Erich Karkoschka, Siddharth Hegde, Paul Helfenstein, Steve Squyres, and our reviewers for helpful discussions and comments. The authors acknowledge support by the Simons Foundation (SCOL # 290357, Kaltenegger), the Carl Sagan Institute, and the NASA/New York Space Grant Consortium (NASA Grant # NNX15AK07H).

AUTHOR DISCLOSURE STATEMENT No competing financial interests exist.

**Table 1.** Parameters for the 10 Solar System bodies from the THN-PSL we used to calculate the calibrated flux and albedos. References used for phase function and albedo. [†]To obtain the proper geometric albedo for this Earthshine observation a factor of 2.38E5 is needed. [*]The Pluto-Charon spectrum is added for completeness.

| Name | Obs. Date | V Mag. | $d$ (AU) | $a_b$ (AU) | $R_b$ (km) | $\alpha$ (deg.) | $\phi(\alpha)$ | Phase Ref. | Albedo Ref. |
|---|---|---|---|---|---|---|---|---|---|
| Ceres | 11/25/08 | 8.40 | 2.405 | 2.558 | 470 | 23 | 0.34(5) | a | a |
| Dione | 5/5/08 | 10.40 | 8.97 | 9.298 | 560 | 6 | 0.88(3) | b | l,v |
| Earth | 11/21/08 | -2.50 | 0.0026[†] | 0.988[†] | 6378 | 70 | 0.5 | c | m,n |
| Jupiter | 5/7/08 | -2.40 | 4.68 | 5.199 | 71492 | 10 | 0.91(8) | d | o |
| Moon | 11/21/08 | -9.30 | 0.0026 | 0.988 | 1738 | 108 | 0.05(1) | e | e |
| Neptune 1 | 5/7/08 | 7.90 | 30.14 | 30.04 | 24766 | 2 | 1 | f | o,p |
| Neptune 2 | 11/20/08 | 7.90 | 30.145 | 30.03 | 24766 | 2 | 1 | f | o,p |
| Neptune 3 | 11/25/08 | 7.90 | 30.23 | 30.03 | 24766 | 2 | 1 | f | o,p |
| Neptune 4 | 11/26/08 | 7.90 | 30.247 | 30.03 | 24766 | 2 | 1 | f | o,p |
| Pluto[*] | 5/11/08 | 15.00 | 30.72 | 31.455 | 1150 | 1 | 1 |  | q |
| Rhea | 11/25/08 | 9.90 | 9.591 | 9.36 | 764 | 6 | 0.87(2) | b | l,v |
| Saturn 1 | 5/5/08 | 1.00 | 8.97 | 9.296 | 60268 | 6 | 0.76(3) | d | o,p |
| Saturn 2 | 11/19/08 | 1.20 | 9.685 | 9.356 | 60268 | 6 | 1 | d | o,p |
| Saturn 3 | 11/19/08 | 1.20 | 9.685 | 9.356 | 60268 | 6 | 0.76(3) | d | o,p |
| Saturn 4 | 11/22/08 | 1.20 | 9.6385 | 9.359 | 60268 | 6 | 0.76(3) | d | o,p |
| Titan 1 | 5/5/08 | 8.40 | 8.97 | 9.3 | 2575 | 6 | 0.98(3) | g | o,p |
| Titan 2 | 5/6/08 | 8.40 | 8.97 | 9.3 | 2575 | 6 | 0.98(3) | g | o,p |
| Titan 3 | 11/24/08 | 8.60 | 9.5947 | 9.364 | 2575 | 6 | 0.98(3) | g | o,p |
| Uranus 1 | 5/11/08 | 5.90 | 20.66 | 20.097 | 25559 | 2 | 1 | f | o,p |
| Uranus 2 | 11/20/08 | 5.80 | 19.742 | 20.097 | 25559 | 3 | 1 | f | o,p |

References for Table 1 and 2: [a](Reddy et al., 2015), [b](Buratti & Veverka, 1984), [c](Goode et al., 2001), [d](Irvine et al., 1968), [e](Lane & Irvine, 1973), [f](Pollack et al., 1986), [g](Tomasko & Smith, 1982), [h](Squyres & Veverka, 1981), [i](Buratti & Veverka, 1983), [j](Simonelli & Veverka, 1984), [k](Mallama et al., 2002), [l](Noll et al., 1997), [m](Kaltenegger et al., 2010), [n](Meadows, 2006), [o](Karkoschka, 1998), [p](Fink & Larson, 1979), [q](Lorenzi et al., 2016; Protopapa et al., 2008) [r](Spencer, 1987), [s](Spencer et al., 1995) [t](Fanale et al., 1974) [u](McCord & Westphal, 1971), [v](Cassini VIMS - NASA PDS), [w](Pollack et al., 1978), [x](Mallama, 2017)

**Table 2.** Data for the Solar System bodies from the THN-PSL dataset that were contaminated based on the shape of their calculated geometric albedo. *Note that the authors state that the Callisto data is contaminated.

| Name | Obs. Date | V Mag. | $d$ (AU) | $a_b$ (AU) | $R_b$ (km) | $\alpha$ (deg.) | $\phi(\alpha)$ | Phase Ref. | Albedo Ref. |
|---|---|---|---|---|---|---|---|---|---|
| Callisto* | 5/5/08 | 6.30 | 4.73 | 5.214 | 2410 | 10 | 0.60(2) | h | r,s |
| Europa | 5/7/08 | 5.60 | 4.69 | 5.195 | 1565 | 10 | 0.88(5) | i,b | i,r,s |
| Ganymede | 11/26/08 | 5.40 | 5.7518 | 5.12066 | 2634 | 8 | 0.80(5) | h | r,s |
| Io | 11/26/08 | 5.80 | 5.7556 | 5.12348 | 1821 | 8 | 0.87(5) | j | s,t |
| Mars | 5/12/08 | 1.30 | 1.68 | 1.6676 | 3397 | 35 | 0.58(5) | d | u,n |
| Mercury | 5/11/08 | 0.00 | 0.99 | 0.37547 | 2440 | 96 | 0.11(5) | d,k | k,x |
| Venus | 11/20/08 | -4.20 | 1.077 | 0.72556 | 6052 | 63 | 0.4(1) | d | n,m |

References for Table 1 and 2: [a](Reddy et al., 2015), [b](Buratti & Veverka, 1984), [c](Goode et al., 2001), [d](Irvine et al., 1968), [e](Lane & Irvine, 1973), [f](Pollack et al., 1986), [g](Tomasko & Smith, 1982), [h](Squyres & Veverka, 1981), [i](Buratti & Veverka, 1983), [j](Simonelli & Veverka, 1984), [k](Mallama et al., 2002), [l](Noll et al., 1997), [m](Kaltenegger et al., 2010), [n](Meadows, 2006), [o](Karkoschka, 1998), [p](Fink & Larson, 1979), [q](Lorenzi et al., 2016; Protopapa et al., 2008) [r](Spencer, 1987), [s](Spencer et al., 1995) [t](Fanale et al., 1974) [u](McCord & Westphal, 1971), [v](Cassini VIMS - NASA PDS), [w](Pollack et al., 1978), [x](Mallama, 2017)

# APPENDIX

**Table A1.** Calculated k values for each band and their average for each observation in the THN-PSL. The CV and albedo (Fig. 3) was used to determine level of reliability of the observation.

| Name | Obs. Date | $k_V$ | $k_R$ | $k_I$ | $k_J$ | $k$ | StDev | CV |
|---|---|---|---|---|---|---|---|---|
| *Uncontaminated (CV < 6% and albedo < 1)* | | | | | | | | |
| Ceres | 11/25/08 | 8.27E-16 | 8.27E-16 | 8.45E-16 | 8.36E-16 | 8.34E-16 | 8.44E-18 | 1.01% |
| Dione | 5/5/08 | 1.34E-16 | 1.34E-16 | 1.35E-16 | 1.34E-16 | 1.34E-16 | 6.75E-19 | 0.50% |
| Earth | 11/21/08 | 1.48E-11 | 1.48E-11 | 1.50E-11 | 1.50E-11 | 1.49E-11 | 1.14E-13 | 0.77% |
| Jupiter | 5/7/08 | 2.17E-11 | 2.16E-11 | 2.08E-11 | 2.19E-11 | 2.15E-11 | 5.03E-13 | 2.34% |
| Moon | 11/21/08 | 1.49E-08 | 1.49E-08 | 1.54E-08 | 1.51E-08 | 1.51E-08 | 2.27E-10 | 1.50% |
| Neptune 1 | 5/7/08 | 5.28E-16 | 5.38E-16 | 4.89E-16 | 5.25E-16 | 5.20E-16 | 2.14E-17 | 4.11% |
| Neptune 2 | 11/20/08 | 4.44E-16 | 4.53E-16 | 4.33E-16 | 4.42E-16 | 4.43E-16 | 8.44E-18 | 1.90% |
| Neptune 3 | 11/25/08 | 2.96E-16 | 3.02E-16 | 2.86E-16 | 2.97E-16 | 2.95E-16 | 6.51E-18 | 2.20% |
| Neptune 4 | 11/26/08 | 7.80E-16 | 7.94E-16 | 7.46E-16 | 7.76E-16 | 7.74E-16 | 2.00E-17 | 2.59% |
| Pluto | 5/11/08 | 2.67E-18 | 2.67E-18 | 2.70E-18 | 2.71E-18 | 2.69E-18 | 2.10E-20 | 0.78% |
| Rhea | 11/25/08 | 2.72E-16 | 2.71E-16 | 2.73E-16 | 2.73E-16 | 2.72E-16 | 1.14E-18 | 0.42% |
| Saturn 1 | 5/5/08 | 8.29E-13 | 8.31E-13 | 7.86E-13 | 8.49E-13 | 8.24E-13 | 2.70E-14 | 3.28% |
| Saturn 2 | 11/19/08 | 1.11E-12 | 1.10E-12 | 1.05E-12 | 1.13E-12 | 1.10E-12 | 3.44E-14 | 3.14% |
| Saturn 3 | 11/19/08 | 1.00E-12 | 9.97E-13 | 9.57E-13 | 1.02E-12 | 9.94E-13 | 2.68E-14 | 2.70% |
| Saturn 4 | 11/22/08 | 7.57E-13 | 7.52E-13 | 7.21E-13 | 7.62E-13 | 7.48E-13 | 1.81E-14 | 2.42% |
| Titan 1 | 5/5/08 | 1.12E-15 | 1.13E-15 | 1.09E-15 | 1.13E-15 | 1.12E-15 | 1.79E-17 | 1.60% |
| Titan 2 | 5/6/08 | 1.74E-15 | 1.72E-15 | 1.70E-15 | 1.76E-15 | 1.73E-15 | 2.47E-17 | 1.43% |
| Titan 3 | 11/24/08 | 1.17E-15 | 1.16E-15 | 1.12E-15 | 1.18E-15 | 1.16E-15 | 2.45E-17 | 2.11% |
| Uranus 1 | 5/11/08 | 3.04E-15 | 3.12E-15 | 2.76E-15 | 3.00E-15 | 2.98E-15 | 1.53E-16 | 5.12% |
| Uranus 2 | 11/20/08 | 3.30E-15 | 3.37E-15 | 3.19E-15 | 3.26E-15 | 3.28E-15 | 7.48E-17 | 2.28% |
| *Contaminated (albedo > 1)* | | | | | | | | |
| Callisto | 5/5/08 | 7.38E-15 | 7.36E-15 | 6.90E-15 | 7.55E-15 | 7.30E-15 | 2.78E-16 | 3.81% |
| Europa | 5/7/08 | 2.47E-14 | 2.46E-14 | 2.58E-14 | 2.48E-14 | 2.49E-14 | 5.57E-16 | 2.24% |
| Ganymede | 11/26/08 | 4.20E-14 | 4.17E-14 | 4.52E-14 | 4.26E-14 | 4.29E-14 | 1.62E-15 | 3.79% |
| Io | 11/26/08 | 1.26E-14 | 1.25E-14 | 1.41E-14 | 1.29E-14 | 1.31E-14 | 7.45E-16 | 5.70% |
| Mars | 5/12/08 | 1.59E-12 | 1.57E-12 | 1.73E-12 | 1.60E-12 | 1.62E-12 | 7.56E-14 | 4.67% |
| Mercury | 5/11/08 | 7.18E-12 | 7.12E-12 | 8.02E-12 | 7.31E-12 | 7.41E-12 | 4.16E-13 | 5.61% |
| Venus | 11/20/08 | 1.40E-10 | 1.39E-10 | 1.43E-10 | 1.39E-10 | 1.40E-10 | 2.08E-12 | 1.49% |
| *Contaminated (CV >14%)* | | | | | | | | |
| Earth | 5/11/08 | 1.84E-11 | 1.79E-11 | 1.62E-11 | 2.26E-11 | 1.88E-11 | 2.71E-12 | 14.43% |
| Moon | 11/21/08 | 2.08E-08 | 1.84E-08 | 2.10E-08 | 3.04E-08 | 2.26E-08 | 5.33E-09 | 23.55% |
| Uranus | 5/7/08 | 3.28E-15 | 3.22E-15 | 2.60E-15 | 7.30E-16 | 2.46E-15 | 1.19E-15 | 48.52% |